\documentclass[12pt,preprint]{aastex}
\usepackage[numberedappendix]{emulateapj5}  % manuscript mode, emulate ApJ

\slugcomment{accepted for ApJ}

\shorttitle{Cluster Heating by Waves}
\shortauthors{Ruszkowski et al.}

\received{2003 October 27}
\begin{document}

\title{Cluster Heating by Viscous Dissipation of Sound Waves}

\author{Mateusz Ruszkowski\altaffilmark{1}}
\affil{JILA, Campus Box 440, 
University of Colorado at Boulder, CO 80309-0440; \email{mr@quixote.colorado.edu}}

\author{Marcus Br\"{u}ggen}
\affil{International University Bremen, Campus Ring 1, 28759 Bremen, Germany; 
\email{m.brueggen@iu-bremen.de}}
\and
\author{Mitchell C. Begelman\altaffilmark{2}}
\affil{JILA, Campus Box 440, University of Colorado at Boulder, CO 80309-0440;
\email{mitch@jila.colorado.edu}}

\altaffiltext{1}{{\it Chandra} Fellow}
\altaffiltext{2}{also at Department of Astrophysical and Planetary Sciences, 
University of Colorado at Boulder}

\begin{abstract}
We simulate the effects of viscous dissipation of waves that are generated
by AGN activity in clusters of galaxies. We demonstrate that the amount of
viscous heating associated with the dissipation of these waves can offset
radiative cooling rates in cooling flow clusters of galaxies. This heating
mechanism leads to spatially distributed and approximately symmetrical
dissipation.  The heating waves reach a given distance from the cluster
center on a timescale shorter than the cooling time.  This means that this
heating mechanism has the potential of quenching cooling flows in a
quasi-stable fashion. Moreover, the heating is gentle as no strong shocks
are present in the simulations. We first investigated whether a single
continuous episode of AGN activity can lead to adequate dissipation to
balance cooling rates.  These simulations demonstrated that, whereas
secondary waves generated by the interaction of the rising bubble with the
intracluster medium are clearly present, viscous heating associated with
the dissipation of these waves is insufficient to balance radiative
cooling. It is only when the central source is intermittent that the
viscous dissipation of waves associated with subsequent episodes of
activity can offset cooling. This suggests that the ripples observed in the
Perseus cluster can be interpreted as being due to the AGN duty cycle,
i.e., they trace AGN activity history. The simulations were performed using
the PPM adaptive mesh refinement code FLASH in two dimensions.

\end{abstract}


\section{Introduction}

Cooling timescales of gas in the central regions of clusters of galaxies
are often much shorter than the Hubble time. Initially, this led to
suggestions that the intracluster medium (ICM) is flowing into the cluster
center at rates of up to 1000 $M_{\odot}/$yr.  However, recent {\it XMM
Newton} and {\it Chandra} observations suggest that the actual inflow rates
are much smaller than expected, and that feedback from active galactic
nuclei (AGN) may play a crucial role in regulating mass accretion rates
(e.g., \citet{fab00,fab02,mc00,bla01,ch02}). The significance of AGN
feedback is supported by the observation that about 70\% of cD galaxies in
cluster centers show evidence for active radio sources \citep{bur90}.  The
advantage of the AGN heating model over other models is that the heating is
supplied near the cluster center where the cooling flow problem is most
severe. For example, AGN heating may explain why the gas temperature, while
declining towards cluster centers, does not drop below about 1 keV
\citep{pe01,pe03,ta01}.\\ 
\indent 
AGN are believed to be intermittent with an intermittency period of
$10^{5}-10^{8}$ yr, much shorter than the Hubble time and shorter than or
comparable to the central cooling time (e.g.,
\citet{maz02,cro03,fab03a,fab03b}). Therefore, one expects that AGN-heated
cooling flows could be stabilized in a time-averaged sense and that
``cooling catastrophes'' could be prevented.  Recent observations of
ripples and weak shocks in the Perseus cluster \citep{fab03a,fab03b} 
and the Virgo cluster \citep{for03}
provide observational support for this idea. \citet{fab03a} were the first
to show that viscous dissipation of these waves is sufficient to offset
radiative cooling in the Perseus cluster.\\
\indent 
Recently, several studies have addressed the problem of AGN heating
of clusters from a numerical perspective. These studies can be divided into
two main categories depending on the parameter regime considered: models in
which the mechanical energy supply to the cluster is momentum driven (e.g.,
\citet{ta93,re01}) and those in which it is buoyancy driven (e.g.,
\citet{ch01,br02,bk02,bru03,qu01}). In this paper we focus on the latter
regime. An alternative idea was proposed by \citet{pr89}, who suggested
that clusters can be heated by dissipation of sound waves generated by
galaxy motions in the cluster. Further support for the idea that viscosity
may play an important role in the intracluster medium comes from the recent
study of density profiles in clusters \citep{hs03}.  The main purpose of
this paper is to demonstrate that clusters can be heated efficiently by
wave dissipation associated with activity of AGN located in their centers.
Although our simulations are two-dimensional and therefore not directly
applicable to real clusters, we argue that the basic result should be
preserved in three dimensions. 

\section{Assumptions of the Model}

\subsection{Initial Conditions}
The intracluster medium is initially assumed to be in hydrostatic
equilibrium in an NFW potential \citep{na95,na97} for which the
gravitational acceleration as a function of the distance from the cluster
center $r$ is given by

\begin{equation}
\mathbf{g}(r)=4\pi G\rho_{\rm crit}\delta_{c}r_c x^{-2}
\left[-\ln (1+x)+\frac{x}{1+x}\right]\mathbf{\hat{r}},
\end{equation}

\noindent
where $r_{c}=100$ kpc is the core radius, $x=r/r_{c}$,
$\mathbf{\hat{r}}=\mathbf{r}/r$, 
$\delta_c = 3.0\times 10^{4}$ is the central overdensity, and
$\rho_{\rm crit}=3H_{o}^{2}/(8\pi G)$ is the critical 
density of the Universe (we assume $H_{o}=75$ km s$^{-1}$ Mpc$^{-1}$).
The initial temperature distribution is given by

\begin{equation}
T(r)=T_{o}\left(1+\frac{r}{r_{o}}\right)^{\beta},
\end{equation}

\noindent
where $T_{o}=3.0$ keV, $r_{o}=10$ kpc and $\beta =0.22$. The temperature at
100 kpc is $5.1$ keV.  The central electron number density is $2.8\times
10^{-2}$ cm$^{-3}$. The electron number density at 100 kpc is approximately
$5.3\times 10^{-3}$ cm$^{-3}$. This corresponds to a central cooling time
of $\sim 1.3\times 10^{9}$ years and a cooling time of $\sim 1.0\times
10^{10}$ years at 100 kpc. These are values typical of a cooling flow
cluster as the central cooling time is much shorter than the Hubble
time. Note that we do not intend to model the details of any particular
cluster even if our studies have been motivated by observations of
Perseus. Since our simulations are two-dimensional, a quantitative
comparison is not possible. For one thing, the wave energy density can be
expected to decay more slowly with radius in two dimensions ($\propto
r^{-1}$) than in three ($\propto r^{-2}$).  Moreover, bubbles expanding
into two dimensions are expected to be larger than equivalent bubbles
expanding three-dimensionally; indeed, for our initial conditions chosen to
approximate those in Perseus, we find that our computed bubbles are larger
than the observed cavities.

We assume that the gas is fully ionized and characterized by
$X=0.75$ and $Y=0.25$, where $X$ and $Y$ are the hydrogen and helium
fractions. The injected gas is characterized by an adiabatic index
$\gamma_{\rm bubble}=4/3$, whereas for the ambient gas we used $\gamma_{\rm
ICM}=5/3$.  Calculations were done in two dimensions for 9 levels of
refinement using the PPM adaptive mesh refinement code FLASH. The size of
the computational domain was (200 kpc)$^2$. Thus, the effective resolution
in our simulations was 2048$^2$ zones, which corresponds to $\sim 0.1$
kpc. We have performed convergence tests and found that neither the
geometry of the bubbles nor that of the waves depends on the adopted
resolution. This is due to the fact that, for the parameters considered,
the Reynolds number corresponding to bubbles and waves is low and also
because numerical dissipation of PPM codes is known to be relatively low.
The effective Reynolds numbers achievable in the simulation are
proportional to the number of grid points across the fluctuation of
interest to the power $n$, where $n=3$ is the order of the numerical
scheme\footnote{See, e.g., \citet{ph03} for the definition of ``the order
of the numerical scheme'', as it is different from the customary definition
of accuracy of a perturbative calculation.}  \citep{syt00,bal96,por94}. We
used a redefined system of units in which all variables apart from
temperature are close to unity and adopted outflow boundary conditions.  We
have also repeated our simulations in two-dimensional cylindrical
coordinates and found that, in spite of unphysical effects near the
symmetry axis, our conclusions remain unaffected.

\subsection{Heating}
We model AGN heating by injecting hot gas into two regions of radius 1 kpc
located 10 kpc to either side of the cluster center.  The energy injection
rate, $L$, for each source and the mass injection rate per unit volume,
$\dot{\rho}$, are both constant. Thus, the energy injection rate per unit
mass $\dot{\epsilon}$ is computed from

\begin{equation}
\dot{\epsilon}=\left(\frac{L}{\rho V}-\epsilon\frac{\dot{\rho}}{\rho}\right),
\end{equation}

\noindent
where $V$ is the volume of one injection region (of radius 1 kpc). We used
$L=1.5\times 10^{43}$ erg s$^{-1}$ and $\dot{\rho}V=0.01$
M$_{\odot}$yr$^{-1}$.  The energy injection is intermittent with an
intermittency period of $3\times 10^{7}$ years, i.e., the source is active
for $1.5\times 10^{7}$ years and dormant for $1.5\times 10^{7}$ years. In
the initial state for each activity episode, the temperature and density
are a hundred times higher and lower, respectively, than the temperature
and density in the initial unperturbed state at the same location.\\
\indent The dissipation of mechanical energy due to viscosity, per unit
mass of the fluid, was calculated from \citep{bat67,shu92}

\begin{equation}
\dot{\epsilon}_{\rm visc}=\frac{2\mu}{\rho}\left(e_{ij}e_{ij}-\frac{1}{3}\Delta^{2}\right),
\end{equation}

\noindent
where $\Delta =e_{ii}$ and

\begin{equation}
e_{ij}=\frac{1}{2}\left(\frac{\partial v_{i}}{\partial x_{j}}+\frac{\partial v_{j}}{\partial x_{i}}\right),
\end{equation}

\noindent
and where $\mu$ is the dynamical coefficient of viscosity. We use the
standard Spitzer viscosity \citep{bra58}, for which $\mu = 1.1\times
10^{-16}T^{5/2}$ g cm$^{-1}$ s$^{-1}$.  As conditions inside the buoyantly
rising bubbles are very uncertain and because we want to focus on energy
dissipation in the ambient ICM, we assume that dissipation occurs only in
the regions surrounding the buoyant gas. To this end we impose a condition
that switches on viscous effects provided that the fraction of the injected
gas in a given cell is much smaller than unity. We point out that the value
of viscosity in the ICM, just as any other transport parameters such as,
e.g., thermal conduction, is highly uncertain, and especially the role of
magnetic fields is unclear.\\ \indent We implemented a fully compressible
version of the viscous velocity diffusion equation in the FLASH code.
Velocity diffusion was simulated by solving the momentum equation

\begin{equation}
\frac{\partial (\rho v_{i})}{\partial t}+
\frac{\partial}{\partial x_{k}}(\rho v_{k}v_{i})+
\frac{\partial P}{\partial x_{i}} = \rho g_{i}+
\frac{\partial\pi_{ik}}{\partial x_{k}} ,
\end{equation}

\noindent
where

\begin{equation}
\pi_{ik}=\frac{\partial}{\partial x_{k}}\left[2\mu 
\left(e_{ik}-\frac{1}{3}\Delta\delta_{ik}\right)\right]
\end{equation}

\noindent
and where all other symbols have their usual meaning. 

\subsection{Cooling}

We switched off radiative cooling because the initial cooling time in the
center is longer than the overall duration of the simulation. However, we
calculate the radiative cooling rates in order to compare them with the
viscous heating rates. For this purpose we use the fit to the cooling
function by \citet{to01}, which is based on detailed calculations by
\citet{sd93}

\begin{equation}
n_{e}^{2}\Lambda =[C_{1}(k_{B}T)^{\alpha}+C_{2}(k_{B}T)^{\beta}+C_{3}]n_{i}n_{e},
\end{equation}

\noindent
where $n_{i}$ is the ion number density and the units for $k_{B}T$ are keV.
For an average metallicity $Z=0.3 Z_{\odot}$, the constants in equation (8)
are $\alpha =-1.7$, $\beta =0.5$, $C_{1}=8.6\times 10^{-3}$,
$C_{2}=5.8\times 10^{-2}$ and $C_{3}=6.4\times 10^{-2}$ and we can
approximate $n_{i}n_{e}=(X+0.5Y)(X+0.25Y)(\rho/m_{p})^{2}$.  The units of
$\Lambda$ are $10^{-22}$ erg cm$^{3}$ $s^{-1}$.

\section{Results and Conclusions}

%\begin{figure*}
%\begin{center}
%\includegraphics[angle=0,width=17cm]{f1.eps}
%%\includegraphics[angle=0,width=17cm]{newf1.eps}
%\caption{Top panels show the evolution of density due to intermittent energy injection near the cluster center. Unit mass density in code
%units corresponds to a real electron number density of 8.75$\times 10^{-3}$
%cm$^{-3}$. Bottom panels show viscous energy dissipation rate. All plots
%are in logarithmic scale. Unit energy dissipation
%rate per unit mass in code units corresponds to 3.24 cm$^2$ s$^-3$.  Snapshots correspond to $5.8\times 10^{7}$, $1.2\times 10^{8}$, $1.8\times 10^{8}$ and $2.3\times 10^{8}$ years, respectively.  The source is periodically active for $1.5\times 10^{7}$ years and dormant for $1.5\times 10^{7}$ years.}
%\end{center}
%\end{figure*}
%
%
%\begin{figure*}
%\begin{center}
%\includegraphics[angle=270,width=10cm]{f2.eps}
%\caption{The ratio of viscous heating to radiative cooling rate as a function of time for a number of concentric annuli around the cluster center.
%The curves that start rising at later times correspond to annuli located further away from the center. The heating-to-cooling ratio was calculated in ten
%rings starting from the first ring at 5 kpc and the remaining rings located in increments of 10 kpc away from the cluster center. Note that the heating
%rate is comparable to the cooling rate. Note also that the curves display a pronounced periodic behavior that  reflects the intermittency of the central source.}
%\end{center}
%\end{figure*}

The top panels in Figure 1 show a time sequence of density maps. One can observe that the gas rises subsonically in the cluster atmosphere and
spreads out laterally. No strong shocks are present in this simulation, which implies that heating is gentle in agreement with {\it Chandra}
observations.  Density waves have maximum amplitudes of up to about 20 to 30 per cent close to the cluster center ($\sim$ 20 kpc) and decrease as the
waves propagate outward.\\ 
\indent
The bottom panels in Figure 1 present the evolution of the viscous dissipation rate. Heating waves generated by subsequent AGN activations are
clearly visible and the energy dissipated in these waves is spatially distributed in a relatively symmetric manner. The timescale for the wave
pattern to reach a particular region is shorter than the local cooling time. Thus, this heating mechanism meets at least one of the basic
requirements for this model to be able to reach a quasi-steady state. We note that the wave fronts propagate at slightly above the sound speed (Mach
number $\sim 1.25$; faster than the buoyantly rising bubbles) as can be seen from Figure 2 by dividing the radius of an annulus by the time it
takes for the wave reach it. \\ 
\indent
It is quite likely that more than one ripple is generated per episode of AGN activity. That is, subsequent outbursts may occur when the bubble has not yet settled down from the previous outburst and it still overpressured, leading to complex time-dependence.  We assumed that the radio source is intermittent on a time scale of
$1.5\times 10^7$ yrs. This simplified assumption on the behavior of the source reproduces two phenomena: (i) the inflation of two well-defined
cavities from the cumulative effects of multiple outbursts and (ii) the production of a number of ripples or density-waves
that propagate radially outward at the speed of sound, as the pressure pulse from each outburst inflates the expanding cavity slightly (see
Fig. 1). The fragmentary, scalloped appearance of the cavities and sound waves is probably overemphasized because the simulation is two-dimensional. The small-scale structure would presumably be suppressed in a three-dimensional model viewed in projection onto the plane of the sky. \\
\indent
Note that the waves disperse as they propagate away from the center. This dispersion is almost entirely due to explicit velocity
diffusion, as tests without this effect have demonstrated. We stress that our use of the Spitzer viscosity is meant to be illustrative and may not accurately  represent momentum transport in the magnetized intracluster medium.  For one thing, magnetic shear stress is likely to dominate over molecular viscosity in the transport of bulk momentum.  This could either enhance or suppress the dissipation of sound wave, and will almost certainly make the dependence of stress on the velocity field more complicated. For another, in this macroscopic form of momentum transport the rate of dissipation (due to reconnection) is nonlocally related to the stress tensor.  Treatment of these effects will require high-resolution magnetohydrodynamical simulations.  Moreover, magnetic fields could introduce effects similar to bulk viscosity, as a result of plasma
microinstabilities.  In our simulations we neglected bulk viscosity, since
it vanishes for an ideal gas. We note that bulk viscosity, if present, could
dissipate waves even more efficiently. Finally, we have neglected the
effects of thermal conduction, which (assuming Spitzer conductivity) could
damp the sound waves more quickly than Spitzer viscosity (since the
conductive dissipation exceeds viscous one by a factor $\sim 10$ under
simplified assumption that waves are plane and linear and that the gas has
constant density and pressure and gravity can be neglected
\citep{landau}). Since conductivity is expected to be suppressed by
magnetic fields, a realistic assessment of whether conduction enhances the
damping rate of sounds waves is beyond the scope of this investigation.\\
\indent 
Recognizing these caveats, in Figure 2 we compute the ratio of the viscous heating rate to the  radiative cooling rate as a function of time, averaged over a series of
concentric annuli around the cluster center. As the waves need more time to reach the gas located further away from the center, the heating rate rises at progressively later times for more distant annuli. Once the first wave has reached a given distance, viscous heating becomes comparable to the cooling rate.  This is consistent with heating rate predictions made by \citet{fab03a}, also assuming Spitzer viscosity.  We also note that dissipating waves of greater initial amplitude in our simulations would give even more heating to offset
cooling.  Interestingly, the average ratio of heating to cooling seems to be relatively stable as a function of time. We have also computed the volume-integrated heating and cooling rates and found that their ratio converges to a value of the order of a few. However, the balance of heating and cooling is not automatic as it depends on the choice of parameters (e.g., AGN power and density gradient in the intracluster medium) and here feedback may play a role.  Note that the curves display a pronounced
periodic behavior. This reflects the intermittency of the central source, with on- and off-states of $1.5\times 10^{7}$ years. This is consistent with the observational estimates based on {\it Chandra} observations of ripples in the Perseus cluster \citep{fab03a,fab03b}. We performed a series of numerical experiments to investigate whether a single AGN outburst can generate waves for which the dissipation rates could offset local radiative cooling rates. These simulations demonstrated that, whereas secondary waves
generated by the interaction of the rising bubble with the surrounding
intracluster medium are clearly present, the viscous heating associated
with a single outburst is insufficient to balance radiative cooling. This
suggests that the ripples observed in the Perseus cluster can be
interpreted as being due to the AGN duty cycle, i.e., they trace AGN
activity history.\\
\indent 
The work done by the expanding cavities on the ambient medium is
limited to a modest fraction of the energy injected by the AGN. If the
cavities are approximately in pressure balance with their surroundings, the
increase in the energy of the ambient gas, $dU = d(PV)/(\gamma - 1)$, is
related to the work done, $dW = P \, dV$, by $dU \simeq dW/ (\gamma -1)$.
The first law of thermodynamics then implies that $dW \simeq {\gamma - 1
\over \gamma} dQ$, where $dQ$ is the heat injected into the cavity.  This
means that, depending on the effective value of $\gamma$ (which can range
between 4/3 and 5/3),  $25-40\%$ of the energy input can be transferred to
the ambient medium (see also, e.g., Churazov et al. 2001).  The fraction of
the input power transferred to the ICM will be larger if the cavities are
overpressured. The fraction of this work that goes into acoustic energy, as
opposed to other types of disturbance (e.g., g-modes or internal waves),
depends on the timescale of pressure fluctuations, as well as detailed
structure of the cavity--ICM interface.  We expect the production of sound
waves to be efficient when the AGN duty cycle is of the same order as the
sound crossing time at the cavity radius, or shorter.  This condition is
satisfied for our chosen duty cycle of $3 \times 10^7$ yr. In addition to
work done by in situ expansion of the cavities, a roughly comparable amount
of energy is transferred to the surrounding medium, in the form of kinetic
and gravitational potential energy, as the cavities rise through the
backgound pressure gradient.  The latter is the generic mechanism appealed
to by Begelman (2001) and Ruszkowski \& Begelman (2002) in their discussion
of ``effervescent heating". Energy injected in this way can also be
converted to heat through viscous dissipation. Our simulations map the
total viscous dissipation rate, and do not distinguish between dissipation
of sound waves and other kinds of motion.  Note, however, that sound waves
have larger propagation speeds than other modes, and therefore should
progressively dominate the energetics at radii well outside the cavities.\\ 
\indent
Although we have devised a specific model in which the viscous
dissipation rate of sound waves roughly balances local radiative cooling,
such a balance may not be a universal property of AGN heating in cluster
cores. The distribution of sound energy dissipation is largely determined
by the radial structure of the model.  The sound dissipation length for a
fixed wavelength $L(\lambda, r)$ for the parameters in our simulations 
decreases from the center to the edge of
the simulated region, mainly due to the decrease in density (Fabian et
al.~2003).  For the parameters chosen in our simulation the characteristic
dissipation length near the outer edge of the grid is of order the size of
the simulation region, implying that the dissipation is spread over a
volume that far exceeds that of the bubbles, and much of the acoustic
energy goes into heating. The steady rate of production of acoustic energy
then leads to a rough balance between heating and cooling, given the
adopted density and temperature profile. These conditions may not be satisfied in all
clusters. As the sound dissipation length is proportional to the square of
the period of the sound waves, more frequent outbursts should lead to more
centrally concentrated damping. However, the dissipation rate does not
depend on the AGN intermittency period as such, since the pressure pulses
generated by the bubbles are likely to be far from sinusoidal and will
contain a wide range of frequencies. The dispersion of the waves as they
propagate suggest that the ``effective" wavelength will increase with $r$,
an effect that will partially counteract the decrease of $L(\lambda, r)$
with $r$. \\
\indent
Where the velocity field has small-scale structure or where the damping
rate is much larger than the Spitzer rate, acoustic waves (as well as
gravity and internal waves) can be dissipated much closer to the sites
where they are generated.  Distributed heating would then occur only after
the bubbles had penetrated most of the cluster. This is the situation
envisaged by Begelman (2001) and Ruszkowski \& Begelman (2002) in the
effervescent heating scenario.  This form of heating may be occurring
concurrently with the large-scale acoustic heating in Perseus, and may
dominate the heating in other clusters (e.g., those with smaller acoustic
energy generation due to the intermittency properties of the central AGN).
Note that viscosity may help the bubbles penetrate to large distances
without excessive mixing. \\
\indent
We stress that our two-dimensional simulations do not accurately
represent the behavior of three-dimensional acoustic heating in several
respects.  In three dimensions the energy flux in the sound waves decays
faster than in the Cartesian two-dimensional case. Thus, our 2D simulation
might give a more radially distributed heating rate ($\propto r^{-1}$)
compared to a 3D calculation (where the wave energy flux scales as $\propto
r^{-2}$).  However, since the dissipation rate per unit mass is inversely proportional to density 
$n$ while the cooling rate per unit mass is $\propto n$, a slight
steepening of the density profile (by $\sim r^{-1/2}$) should compensate
for the extra power of $r^{-1}$ in the energy flux.  
Note that the scaling of heating and cooling with density does not
necessarily imply instability. The densest central regions, which cool the
fastest, are in fact heated more effectively because the
velocity fluctuations are stronger in the cluster core. The amount of
energy injected to the cluster should also be regulated by the central
cooling rate. That is,  increased cooling rate should lead to more
accretion onto the central AGN.  Accretion of gas onto the center   would
then cause AGN outbursts leading to a reduced central cooling rate.
If acoustic heating is truly able to stabilize radiative cooling, then the
density and the luminosity of the central AGN should adjust automatically,
as it was shown to do in the 1D ZEUS simulations of effervescent heating by
Ruszkowski \& Begelman (2002).
We also expect energetically ``equivalent" bubbles to grow to more rapidly
in 2D than in 3D. Thus it is not surprising that the bubbles in our 2D
simulations are larger than the X-ray holes at the center of the Perseus
Cluster, despite our attempt to roughly match conditions.  We decided to
carry out two-dimensional simulations first because they are
computationally much less demanding and allow us to explore a wider range
of parameters.  However, we are planning to report a limited set of 3D
simulations separately.  Preliminary results suggest that  main conclusions
drawn from the three-dimensional simulations are consistent with those
obtained from 2D simulations.\\
\indent  
In summary, we have demonstrated that viscous heating by an
intermittent AGN located at the center of a cooling flow cluster can
balance radiative cooling and, thus, quench the cooling flow.  Energy is
transferred to the gas by viscous dissipation of waves produced by
intermittent AGN activity with a duty cycle much shorter than the cooling
time. In the proposed heating mechanism, heating is gentle,
spatially-distributed in a symmetric fashion and delivered to the gas
located within the cooling radius faster than the cooling timescale.  In
this first attempt to simulate the effects of dissipation of waves in the
ICM, we have assumed Spitzer viscosity, but we have to concede that the
value of viscosity in the ICM is poorly constrained. Nevertheless, our
results show that this heating mechanism is broadly consistent with the
assumptions of the effervescent heating model \citep{beg01,rus02}, in which
dissipation of waves plays an important role \citep{beg03}, and can be
applied to recently reported observations of ripples in the Perseus
\citep{fab03a,fab03b} and Virgo \citep{for03} clusters.

\acknowledgments

We thank the anonymous referee for many useful comments and Phil Armitage
for his words of wisdom.   In particular, the anonymous referee is
acknowledged for his/her contribution to the discussion of the conversion
of the input energy to the energy in the ICM (paragraph 6 in section 3).
We are grateful to Andy Fabian and Chris Reynolds for their comments, which
improved the paper.  We also thank Peter Ruprecht and Mark Tamisiea for
technical support.  The software used in this work was in part developed by
the DOE-supported ASCI/Alliance Center for Astrophysical Thermonuclear
Flashes at the University of Chicago. We acknowledge support from the
W. M. Keck Foundation, which purchased the JILA 74-processor Keck Cluster.
Some of the calculations presented in this work were performed at National
Center for Supercomputing Applications at the University of Illinois at
Urbana-Champaign, which is funded through the PACI Program at the National
Science Foundation.  Support for this work was provided by National Science
Foundation grant AST-0307502 and the National Aeronautics and Space
Administration through {\it Chandra} Fellowship Award Number PF3-40029
issued by the Chandra X-ray Observatory Center, which is operated by the
Smithsonian Astrophysical Observatory for and on behalf of the National
Aeronautics and Space Administration under contract NAS8-39073.

\clearpage

\begin{figure}
\plotone{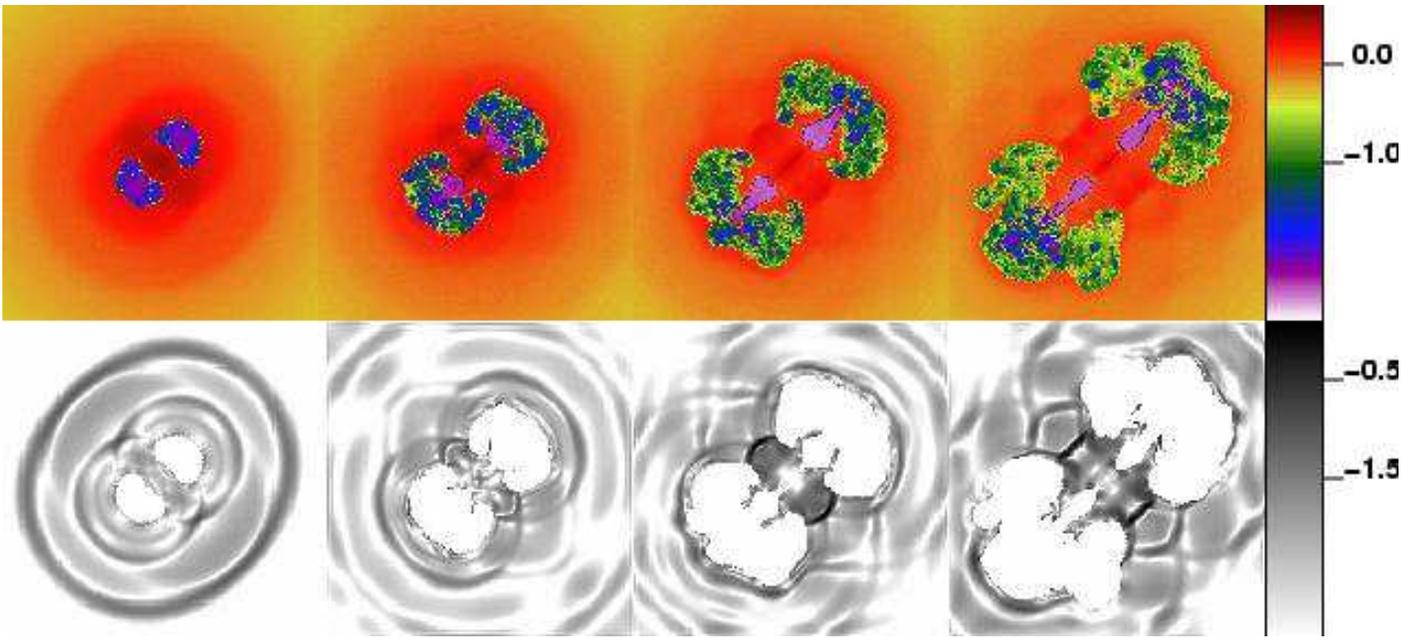}
\caption{Top panels show the evolution of density due to intermittent
energy injection near the cluster center. Unit mass density in code
units corresponds to a real electron number density of 8.75$\times 10^{-3}$ cm$^{-3}$.
Bottom panels show viscous energy dissipation rate. 
All plots are in logarithmic scale. Unit energy dissipation
rate per unit mass in code units corresponds to 3.24 cm$^2$ s$^-3$.
Snapshots correspond to $5.8\times 10^{7}$, $1.2\times 10^{8}$, $1.8\times 10^{8}$
and $2.3\times 10^{8}$ years,
respectively. The source is periodically active for $1.5\times 10^{7}$
years and dormant for $1.5\times 10^{7}$ years.
\label{fig1}}
\end{figure}
\clearpage 
\begin{figure}
%\plotone{f2.eps}
\includegraphics[angle=270]{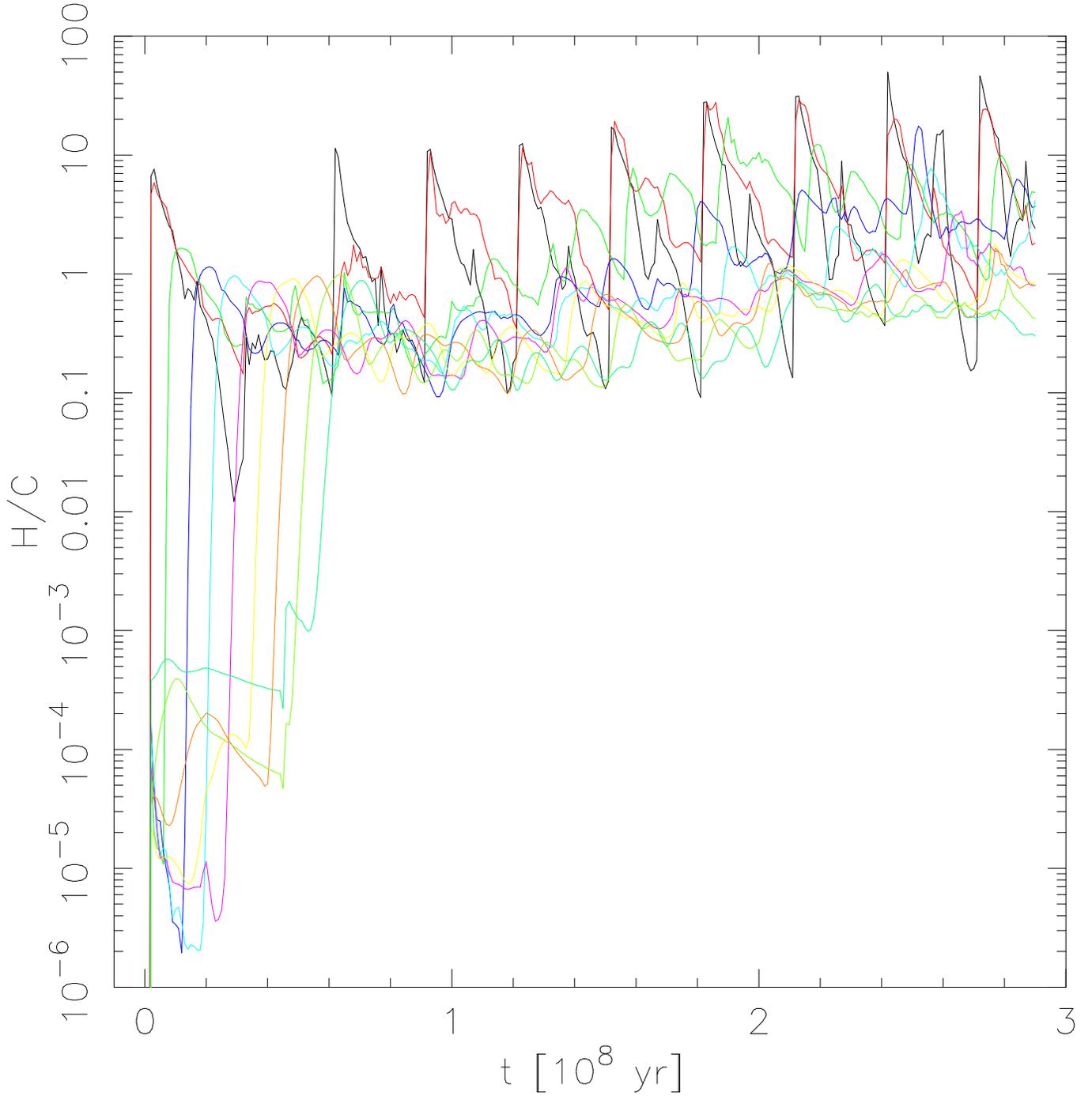}
\caption{The ratio of viscous heating to radiative cooling rate as a
function of time for a number of concentric annuli around the cluster center.
The curves that start rising at later times correspond to annuli located further
away from the center. The heating-to-cooling ratio was calculated in ten
rings starting from the first ring at 5 kpc and the remaining rings
located in increments of
10 kpc away from the cluster center. Note that the heating rate is
comparable to the cooling rate.
Note also that the curves display a pronounced periodic behavior that 
reflects the intermittency of the central source.
\label{fig2}}
\end{figure}

\end{document}